# Scattering from Colloid-Polymer Conjugates with Excluded Volume Effect

Xin Li,[†] Christopher N. Lam,[§] Luis E. Sánchez-Diáz,[†] Gregory S. Smith,[†] Bradley D. Olsen,[*,§] and Wei-Ren Chen[*,†]

[†]Biology and Soft Matter Division, Oak Ridge National Laboratory, Oak Ridge, Tennessee 37831, United States

[§]Department of Chemical Engineering, Massachusetts Institute of Technology, Cambridge, MA 02139, United States

**ABSTRACT:** In this work we present scattering functions of conjugates consisting of a colloid particle and a self-avoiding polymer chain. This model is directly derived from the two point correlation function with the inclusion of excluded volume effects. The dependence of the calculated scattering function on the geometric shapes of the colloid and polymer stiffness is investigated. In comparison to existing experimental results, our model is found to be able to describe the scattering signature of the colloid-polymer conjugates and provide additional conformational information. This model explicitly elucidates the link between the global conformation of a conjugate and the microstructure of its constituent components.

One grand challenge and opportunity in nanoscience is to harness the new phenomena, properties, and functions that emerge by *de novo* design of materials. In this pursuit, polymers offer many advantages because of their chemical diversity and capacity. Under suitable environments or stimuli, they can self-organize into stable structures on a hierarchy of length scales. For example, diblock copolymers spontaneously form a variety of sophisticated, ordered structures whose shape and dimensions can be controlled by tuning the molecular weight and composition of the constituent polymers.[1]

The driving force of this self-assembly phenomenon has been recognized as the minimization of the free energy. In recent years there has been much interest in developing multicomponent building blocks, such as colloid-polymer conjugates, to create programmed, novel functional materials which are thermodynamically favorable on both micro- and mesoscopic length scales.[2-4] The manipulation and synthesis of these heterogeneous building blocks open up unique opportunities to create new material combinations that hold promise for circumventing classic self-organized materials performance trade-offs.[5-14]

To predict and control the self-assembly of these various colloid-polymer amphiphiles, it is critical to elucidate the link between the monomer microstructure and self-assembled phases in the bulk. To this end, small angle scattering techniques[15] have been commonly used to investigate the conformation and positioning of a single colloid-polymer conjugate in these composites[16-29]. Various models of scattering functions have been proposed to extract quantitative structural characteristics from experiment. A composite power law approach has been used to separately obtain the qualitative structural features of colloid-polymer conjugates at different length scales reflected by the characteristic variations of scattering intensities at corresponding $Q$ regimes.[18,19] Global scattering functions which provide unified conformational descriptions have also been proposed. A core-shell model was used to describe the scattering behavior of the systems in which the polymer chains as well as the associated solvent are treated as a homogeneous outer layer[20,21]. Nevertheless, it is clear that this simple model is unable to describe the scattering features originating from the microscopic structure. Alternatively, a mesoscopic approach which treats the polymer in the conjugates as ideal Gaussian chains has been proposed[22,23]. However, in this model excluded volume effects within the polymer chains are not incorporated, and the correlation between a polymer chain and the colloid are calculated based on a phenomenological coarse-grain assumption that the polymer follows Gaussian statistics, with its center of mass located at a certain distance from the colloid. Therefore, it is expected that this model is unable to address the conformational features quantitatively for systems under good solvent conditions.[30] To date viable scattering functions for general colloid-polymer conjugates remain unavailable. This challenge provides the main motivation of this study.

In this letter a small angle scattering function of a colloid-polymer conjugate with explicit incorporation of excluded volume effects is derived. Our model indeed is able to describe the experimental scattering features of colloid-polymer conjugates under good solvent conditions.

The conformation of a colloid-polymer conjugate can be expressed using the two-point spatial correlation function, often called the form factor

$$P_{conjugate}(Q) = \left\langle \sum_{i,j} \exp\left[-i\vec{Q} \cdot (\vec{r}_i - \vec{r}_j)\right] \right\rangle_{\vec{Q}} \quad (1)$$

where the brackets $\langle \rangle$ represent the angular average over $\vec{Q}$, and $\vec{r}_i$ and $\vec{r}_j$ represent the position of two scattering points.[15] For a colloid-polymer conjugate the positions $\vec{r}_i$ and $\vec{r}_j$ in Eqn. (1) run through the whole system. Based on the physical locations of $\vec{r}_i$ and $\vec{r}_j$, $P_{conjugate}(Q)$ can be further expressed as

$$P_{conjugate}(Q) = P_{colloid}(Q) + P_{polymer}(Q) + P_{colloid-polymer}(Q)$$

$$= \left\langle \sum_{i,j} \rho_i \rho_j \exp\left[-i\vec{Q} \cdot (\vec{r}_i - \vec{r}_j)\right] \right\rangle_{\vec{Q},colloid}$$

$$+ \left\langle \sum_{i,j} b_i b_j \delta(\vec{r}_i) \delta(\vec{r}_j) \exp\left[-i\vec{Q} \cdot (\vec{r}_i - \vec{r}_j)\right] \right\rangle_{\vec{Q},polymer}$$

$$+ \left\langle \sum_{i,j} b_i \delta(\vec{r}_i) \rho_j \exp\left[-i\vec{Q} \cdot (\vec{r}_i - \vec{r}_j)\right] \right\rangle_{\vec{Q},colloid-polymer}$$

$$(2)$$

In Eqn. (2) $P_{colloid}(Q)$, $P_{polymer}(Q)$ and $P_{colloid-polymer}(Q)$ represent the scattering contributions respectively from the intra-colloid correlation, the intra-polymer correlation, and the colloid-polymer cross correlations, respectively. $\rho_{i,j}$ gives the scattering length density (SLD) at position $\vec{r}_{i,j}$ in the colloid particle, $b_{i,j}$ is the total bound coherent scattering length of a Kuhn segment with its center at position $\vec{r}_{i,j}$ in the polymer chain.

The contributions from the first and second terms of Eqn. (2) can be written as

$$P_{colloid}(Q) = (\rho V)^2 P_{norm,colloid}(Q) \quad (3)$$

$$P_{polymer}(Q) = \left(b\frac{L}{a_K}\right)^2 P_{WLC}(Q, L, a_K, R_{CS}) \quad (4)$$

where $\rho_{i,j}$ and $b_{i,j}$ reduce to $\rho$ and $b$ respectively with the assumption that the colloid and polymer are both homogeneous. Here $V$ denotes the volume of a colloid particle, $L$ the contour length of the polymer chain, $a_K$ the Kuhn segment length, and $R_{CS}$ the radius of the cross section of Kuhn segments. The pre-factors $(\rho V)^2$ and $(bL/a_K)^2$ give the scattering ability of the colloid and polymer on an absolute intensity scale. $P_{norm,colloid}(Q)$ represents the normalized form factor for the colloid. Various mathematical forms have been derived for particles with different geometric shapes.[15] $P_{WLC}(Q,L,a_K,R_{CS})$ represents the scattering function for a polymer chain with excluded volume effects whose phenomenological expressions have been identified previously.[31,32] The third term on the right hand side of Eqn. (2) represents the cross correlation between the colloid and polymer. It can be further expressed as

$$P_{colloid-polymer}(Q)$$
$$= \left\langle \sum_{i,j} b_i \delta(\vec{r}_i) \rho_j \exp\left[-i\vec{Q} \cdot (\vec{r}_i - \vec{r}_j)\right] \right\rangle_{\vec{Q},colloid-polymer}$$
$$= \left\langle b_i \rho_j \int_L d^3\vec{r}_i \int_V d^3\vec{r}_j \exp\left[-i\vec{Q} \cdot (\vec{r}_i - \vec{r}_j)\right] \right\rangle_{\vec{Q},colloid-polymer} \quad (5)$$
$$= 2b\rho \int_L d^3\vec{r}_i \left\langle \int_V d^3\vec{r}' \exp\left[-i\vec{Q} \cdot \vec{r}'\right] \right\rangle_{\vec{Q},colloid-polymer}$$
$$= 2b\rho \int_L d^3\vec{r}_i \left\langle \int_V d^3\vec{r} \exp\left[-i\vec{Q} \cdot (\vec{r}+\vec{t})\right] \right\rangle_{\vec{Q},colloid-polymer}$$
$$= 2b\rho \int_L d^3\vec{r}_i \left\langle \exp(-i\vec{Q} \cdot \vec{t}) \int_V d^3\vec{r} \exp(-i\vec{Q} \cdot \vec{r}) \right\rangle_{\vec{Q},colloid-polymer}$$
$$= 2b\rho \int_L d^3\vec{r}_i \left\langle \exp(-i\vec{Q} \cdot \vec{t}) F_{colloid}(\vec{Q}) \right\rangle_{\vec{Q},colloid-polymer}$$

where $\vec{r}'$ is the relative difference vector between the positions $\vec{r}_i$ and $\vec{r}_j$. It breaks into the sum of $\vec{t}$ and $\vec{r}$, where $\vec{t}$ is the relative vector from a point on the polymer to the geometric center of the colloid, and $\vec{r}$ is the relative coordinate within the colloid particle, as shown in Figure 1. A factor 2 is assigned to take into account the fact that the two sampling points can be located at the polymer or the colloid with the same possibility. The volume integral

$$F_{colloid}(\vec{Q}) = \int_V d^3\vec{r} \exp(-i\vec{Q} \cdot \vec{r}) \quad (6)$$

gives the Q-dependent scattering amplitude of the colloid particle. $\vec{r}_i$ represents the center of a Kuhn segment which runs through the whole contour length $L$. The relative vector $\vec{t}$ can be calculated based on the end-to-end distance distribution function $p(n,r)$ for a partial chain of $n$ Kuhn segments at a radial distance $r$ and a selected azimuthal angle $\theta$. For an ideal Gaussian chain, $p(n,r)$ follows the Gaussian distribution. As a result, the scattering function takes the expression of the Debye function[33]. Since the excluded volume effect is not incorporated, it is known that the Debye function underestimates the spatial correlation, especially at the high Q regime relevant to the intra-chain correlation.[30] In our study, the mathematical expression of $p(n,r)$ proposed by des Cloizeaux is chosen because of its numerical accuracy as demonstrated by computer simulation with the inclusion of the intra-chain excluded volume effect.[34,35] The excluded volume effect between the polymer and the colloid is considered with exclusion of the solid angle covered by the colloid in $p(n,r)$. The cross correlation between the polymer and colloid can be evaluated numerically.

For a spherical colloid, the scattering amplitude takes the following expression

$$F_{sphere}(Q) = \frac{3V[\sin(QR) - QR\cos(QR)]}{(QR)^3} \quad (7)$$

where $R$ is the sphere radius and $V$ the volume ($V = \frac{4\pi}{3}R^3$). Since $F_{sphere}(Q)$ is an angular independent real number, Eqn. (5) can be further simplified to

$$P_{sphere-polymer}(Q)$$
$$= 2b\rho \int_L d^3\vec{r}_i \frac{3V[\sin(QR) - QR\cos(QR)]}{(QR)^3} \langle \exp(-i\vec{Q}\cdot\vec{t}) \rangle_{\vec{Q},sphere-polymer}$$
$$= 2b\rho \int_L d^3\vec{r}_i \frac{3V[\sin(QR) - QR\cos(QR)]}{(QR)^3} \frac{\sin(Qt)}{Qt}$$
$$= 4\pi b\rho \int_0^{L/a_K} dn \int_0^{na_K} dr \int_0^{\pi} d\theta \sin(\theta) p(n,r) \frac{3V[\sin(QR) - QR\cos(QR)]}{(QR)^3} \frac{\sin(Qt)}{Qt}$$
(8)

where $\frac{\sin(Qt)}{Qt}$ results from the angular average of the kernel function $exp(-i\vec{Q}\cdot\vec{t})$. In Eqn. (8) the integral is explicitly expressed using the end-to-end distance distribution function $p(n,r)$[34].

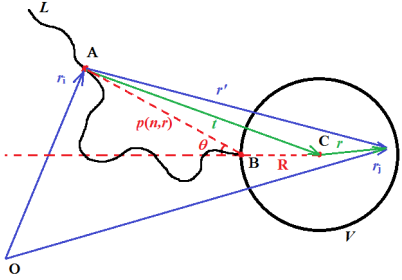

Figure 1. Schematic representation of a conjugate with a homogeneous spherical colloid (volume $V$, centered at $C$) and a wormlike chain (contour length $L$) attached to its surface at point $B$. A partial chain with $n$ monomers follows the end-to-end distance distribution function $p(n,r)$. The red dashed lines indicate the azimuthal angle $\theta$ of the end of the partial chain to the axis defined by the attaching point and the center of the colloid. The colloid-chain cross term in the two-point correlation function (vector $r_i$ and $r_j$) can be calculated numerically via the integral over $n$ and $\theta$.

In Figure 2 we present the form factor $P(Q)$ calculated from Eqns. (2)-(8) for the hard sphere-polymer conjugates. The conformational parameters, $R$, $L$, $a_K$, and $R_{CS}$ used in the calculation are 20 Å, 2667 Å, 10 Å, and 2 Å, respectively. The SLD of the sphere and polymer are both chosen to be 1 so that the total scattering length of the sphere and polymer are equal. The cross contribution between the colloid and the polymer has negative values in the oscillations at some $Q$ values because of the phase interference in the spatial correlation. Therefore, for the sake of clarity, the $P(Q)$'s are shown on a linear scale for the y-axis. Since the total scattering lengths of the colloid and the polymer are the same, the total scattering abilities of the intra-colloid (dashed line) and intra-chain (dotted line) correlations are identical. Therefore, their zero-angle intensities are identical. Since the polymer chain is more loosely packed than the colloid, it spreads over a larger spatial distance, and its Guinier region locates in lower $Q$ range.[15] In the medium $Q$ region ($Q$=0.05~0.2 Å$^{-1}$), the scattering function is dominated by the contribution from the colloid particle. On the other hand, in the high $Q$ region ($Q$>1 Å$^{-1}$), $P(Q)$ is dominated by the intra-polymer correlation due to its slower decay. It should be noted that the summation of the three terms in Eqn. (2) gives the extra concave shape of $P(Q)$ close to the position where the cross correlation and the intra-colloid correlation are equal, for example, $Q$=0.02~0.05 Å$^{-1}$ in Figure 2. This is the specific feature of the scattering profile for colloid-polymer conjugates, which cannot be described using the previous approaches.[16-29] It is important to point out the additional insightful information provided by our calculation. For example, existing scattering spectra of colloid-polymer conjugates[25-29], with similar conformational parameters used in the calculations presented in Fig. 2 (b), are seen to be characterized by an upturn in the $Q$ range of 0.01~0.05 Å$^{-1}$. As demonstrated by Eqn. (5), because of the scattering contribution from the colloid-polymer, the cross correlation is twice that of the intra-colloid and intra-chain correlations; therefore, the zero angle intensity of the whole system is dominated by its contribution (dashed-dotted line). Our calculations show that the experimentally observed upturn could result from the intra-conjugate cross correlation between the colloid and the polymer, instead of the formation of inter-conjugate aggregates or other large length scale heterogeneous structures.

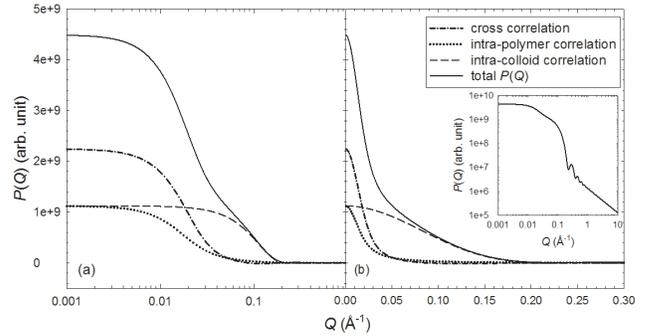

Figure 2. $P(Q)$ of a sphere-polymer conjugate calculated following Eqns. (2)-(8) and shown in (a) semi-logarithm and (b) linear scales. The solid line is the total scattering of $P(Q)$, including the contributions from intra-polymer correlation (dotted), intra-colloid correlation (dash) and the cross correlation between the colloid and polymer (dash-dot). The inset shows the total $P(Q)$ plotted on a logarithm scale.

The calculation of colloid-polymer conjugates can be further extended to investigate colloids with different geometric shapes such as ellipsoids and cylinders, corresponding to the various shapes of proteins or peptides in practical applications. The scattering functions of conjugates consisting of a sphere, ellipsoid, or cylinder with a polymer are calculated via Eqns. (2)-(6). The results are given in panel (a), (b) and (c) of Figure 3, respectively. Their schematic representations are given in the inset of each panel and the comparison of their total scattering functions is given in panel (d). The parameters used in the calculations are chosen as: sphere radius 20 Å, ellipsoid long axis 40 Å, short axes 14.14 Å, cylinder radius 5 Å, cylinder length 320 Å, and the parameters for the polymer are the same as in Figure 2. The volume and total scattering length for the three different colloids are set to be identical. One can notice that from sphere to ellipsoid to cylinder, the colloid is elongated to the axis direction and contracted in the perpendicular direction. As a result, the scattering profile due to the colloid (dashed curve in panels (a)-(c)) extends towards both the low Q and high Q directions. Consequently, as shown in panel (d) the concave region in the total $P(Q)$ near Q=0.03 Å$^{-1}$ becomes progressively pronounced upon increasing the aspect ratio. The oscillation in the calculated total $P(Q)$ of a conjugate also exhibits a strong dependence on the geometric shape of its constituent colloid. For example, due to the large aspect ratio of the cylinder, two sets of oscillations at Q=0.006~0.1 Å$^{-1}$ and 0.5~2 Å$^{-1}$ originating from the correlation along the axis and radius directions, respectively, are observed in our calculations given in panels (c) and (d).

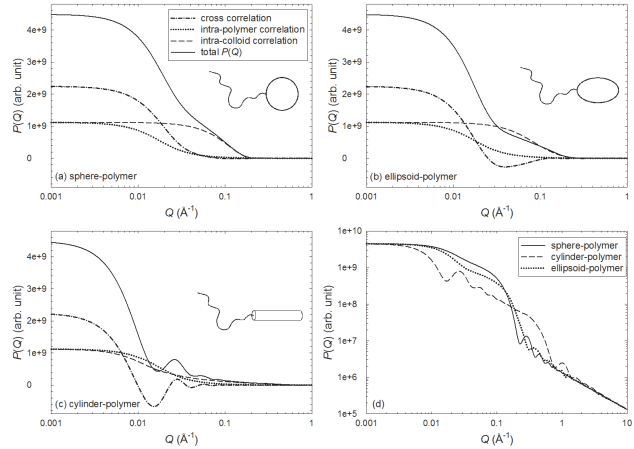

Figure 3. $P(Q)$s of sphere-polymer (a), ellipsoid-polymer (b) and cylinder-polymer (c) conjugates, with intra-colloid (dashed), intra-polymer (dotted), cross (dash-dot), and total (solid) correlations shown separately. The comparison of the total $P(Q)$s for these three cases are presented in panel (d) on a logarithm scale.

Many different chemical and physical stimuli can modulate the conformation of polymer chains. Therefore, it is worth studying the dependence of the scattering function on the stiffness of the polymer chain. For a flexible polymer chain, the ratio $N=L/a_K$ characterizes the stiffness of the polymer chain. The polymer with a larger value of N is characterized by more flexible conformations[33]. From panel (a) to (c) of Figure 4, the Kuhn length decreases from $a_K=R/2$ to $a_K=R/32$, which reflects a deteriorating solvent condition. The intra-polymer correlation shifts towards high Q due to a smaller Kuhn length and the contraction of the global size of the polymer. As a result, the aforementioned concave region in the $P(Q)$ curve becomes less discernible. In the high Q range (Q>0.2 Å$^{-1}$) the intra-polymer scattering contribution (dotted line in panels (a)-(c)) weighs more in the total $P(Q)$, and the oscillations from the intra-colloid correlation is progressively masked by the intra-polymer correlation (panel (d)). As demonstrated in panel (d), $P(Q)$ decays faster in the Porod region (Q>0.8 Å$^{-1}$) with a smaller $a_K$ due to the more compact polymer conformation. Meanwhile, the first minimum of the oscillations (Q=0.15~1 Å$^{-1}$) is seen to shift towards low Q, indicating the size of the colloid indeed increases due to the collapsed polymer attached onto its surface.

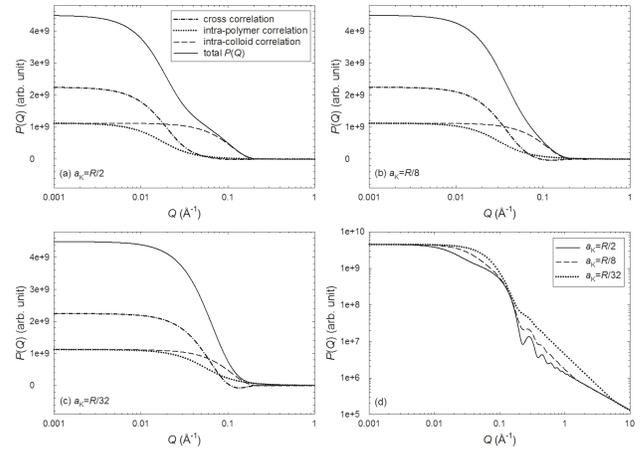

Figure 4. $P(Q)$ of a sphere-polymer conjugate calculated as a function of the Kuhn length $a_K$. The contour length L is fixed at 2667 Å and the sphere radius at 20 Å. The Kuhn lengths are $a_K$=10 Å (a), 2.5 Å (b) and 0.625 Å (c), and their $P(Q)$s are compared in panel (d).

An important question regarding the conformation of protein-polymer conjugates is whether the polymer chain wraps around the protein to create a shield, named the shroud model, or exists as a relatively unperturbed flexible chain attached to the protein, named the dumbbell model.[16] In the extreme case of a shroud model, one can assume that the polymer chain fully collapses on the surface of the colloid, which gives a sphere twice the volume of the spherical colloid in the conjugate. The form factor of this shroud limit is shown in Figure 5 as the dashed line and compared to the sphere-polymer conjugate form factor as the solid line. In SANS measurements, the oscillations in the high Q region (Q>0.1 Å$^{-1}$) may be smeared by the instrument resolution function, but the decay of its intensity can certainly be used to deduce the conformational characteristics of a colloid-polymer conjugate in solution. The shroud model decays much faster, closer to a power law of -4, than the dumbbell model. The concave section of $P(Q)$ at Q=0.03 Å$^{-1}$, originating from the cross correlation, is another feature which can be used for the characterization of the conjugate conformation.

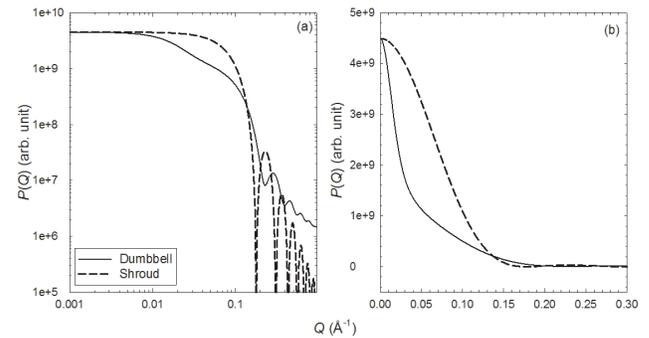

Figure 5. $P(Q)$s of the dumbbell (solid) and shroud models (dashed) of a sphere-polymer conjugate, shown on a logarithm (a) and linear (b) scales. In the shroud

model the polymer is assumed to be fully wrapped around the spherical colloid.

It should be pointed out that all the calculations above are performed under the condition that the total scattering lengths of the colloid and the polymer are equal. In cases where they are not comparable, the scattering profile will be dominated by the one with a much larger scattering length, which renders much stronger scattering ability. Nevertheless, it can be conveniently incorporated into our model once the chemical compositions of materials are specified.

To summarize, we present a theoretical model for the scattering function of colloid-polymer conjugates. This model is directly derived from the two-point spatial correlation function with the incorporation of the excluded volume effect with valid approximations. The scattering functions calculated from this model are able to describe the experimentally observed scattering features of the colloid-polymer conjugates and provide additional insightful interpretations which are not intuitively obvious. This model facilitates the quantitative conformational investigation of colloid-polymer conjugates using scattering.

## AUTHOR INFORMATION

**Corresponding Author**

*E-mail: chenw@ornl.gov; bdolsen@mit.edu

## ACKNOWLEDGMENT

This work was supported by the U.S. Department of Energy, Office of Basic Energy Sciences, Materials Sciences and Engineering Division, and the Scientific User Facility Division.

SYNOPSIS TOC (Word Style "SN_Synopsis_TOC"). If you are submitting your paper to a journal that requires a synopsis graphic and/or synopsis paragraph, see the Instructions for Authors on the journal's homepage for a description of what needs to be provided and for the size requirements of the artwork.

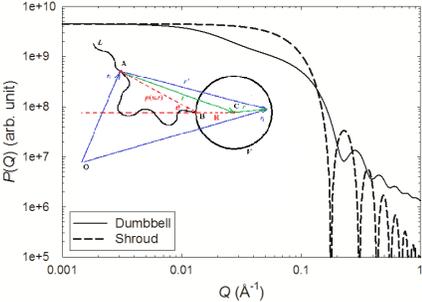